\documentclass[ superscriptaddress, 
                notitlepage, 
                aps,
                prb,
                reprint] 
                {revtex4-1}
\usepackage{hyperref}
\usepackage{color,soul}
\usepackage{amsmath}
\usepackage{amssymb}
\usepackage[caption=false]{subfig}
\usepackage{graphicx}
\usepackage{array}
\usepackage[normalem]{ulem}

\usepackage{xspace}
\newcommand{\Z}{\ensuremath{\mathbb Z}\xspace}
\renewcommand{\S}{\ensuremath{\vec{\mathbb S}}\xspace}
\newcommand{\pointsym}{\ensuremath{\mathcal P}\xspace}
\newcommand{\transsym}{\ensuremath{\mathcal T}\xspace}
\newcommand{\vecrep}{\ensuremath{V}\xspace}
\newcommand{\dblock}{\ensuremath{d_b}}
\newcommand{\ferro}{\ensuremath{\text{ferro}}}
\newcommand{\afm}{\ensuremath{\text{afm}}\xspace}
\usepackage{dsfont}
\newcommand{\identity}{\ensuremath{\mathds{1}}\xspace}

\usepackage{tikz}
\usepackage{pgfplots}
\usetikzlibrary{positioning}
\usetikzlibrary{fit}
\usetikzlibrary{backgrounds}
\usetikzlibrary{calc}
\usetikzlibrary{shapes}
\usetikzlibrary{mindmap}
\usetikzlibrary{arrows}
\usetikzlibrary{decorations.text}
\definecolor{ColorKnown}{HTML}      {A5C8E1}
\definecolor{ColorConsequence}{HTML}{EEA8A9}
\definecolor{ColorDecision}{HTML}   {FFCB9E}
\definecolor{ColorNew}{HTML}        {AAD9AA}
\definecolor{ColorRed}{HTML}        {FF0000}
\definecolor{ColorMagField}{HTML}   {1F77B4}
\newcommand*\blockmatrix[1]{
\tikz[baseline=(char.base)]{\node[shape=rectangle,draw,inner sep=2pt] (char) {#1};}}
\newcommand{\cmt}
{
    Condensed Matter Theory Group,
    Paul Scherrer Institute,
    5232 Villigen PSI,
    Switzerland
}
\newcommand{\theoethz}
{
    Institute for Theoretical Physics,
    ETH Zurich,
    8093 Zurich,
    Switzerland
}
\newcommand{\mesosysethz}
{
    Laboratory for Mesoscopic Systems,
    Department of Materials,
    ETH Zurich,
    8093 Zurich,
    Switzerland
}
\newcommand{\mesosyspsi}
{
    Laboratory for Multiscale Materials Experiments,
    Paul Scherrer Institute,
    5232 Villigen PSI,
    Switzerland
}
\newcommand{\unigeneva}{
    Department of Theoretical Physics,
    University of Geneva,
    1211 Geneva,
    Switzerland
}
\date{\today}

\begin{document}


\title{
    Continuous ground-state degeneracy of classical dipoles on regular lattices
}
\author{Dominik \surname{Schildknecht}}
\email{dominik.schildknecht@gmail.com}
\affiliation{\cmt}
\affiliation{\mesosysethz}
\affiliation{\mesosyspsi}
\author{Michael \surname{Sch\"utt}}
\affiliation{\cmt}
\affiliation{\unigeneva}
\affiliation{\theoethz}
\author{Laura J. \surname{Heyderman}}
\affiliation{\mesosysethz}
\affiliation{\mesosyspsi}
\author{Peter M. \surname{Derlet}}
\email{peter.derlet@psi.ch}
\affiliation{\cmt}
\begin{abstract}
    Dipolar interactions are crucial in the modeling of many complex magnetic systems, such as the pyrochlores and artificial spin systems. Remarkably, many classical dipolar-coupled spin systems exhibit a continuous ground-state degeneracy, which is unexpected as the Hamiltonian does not possess a continuous symmetry. In this paper, we explain how such a finite point symmetry leads to a continuous ground-state degeneracy of specific classical dipolar-coupled systems. This work therefore provides new insight into the theory of classical dipolar-coupled spin systems and opens the way to understand more complex dipolar-coupled systems.
\end{abstract}
\maketitle
\section{Introduction}
In the early 20th century, adiabatic demagnetization associated with the magnetocaloric effect~\cite{Warburg1881} was exploited to reach temperatures below 1~K, in particular in the paramagnetic salts~\cite{Giauque1927,Giauque1935}. The magnetic order limits the coldest temperatures achievable with this method~\cite{Kuerti1933,DeHaas1934,VanVleck1937,Sauer1940}, which called for a better understanding of the ordered states in such systems. The difficult problem of the ground state determination in dipolar-coupled spin systems was, however, not successfully tackled until the pioneering work of Luttinger and Tisza~\cite{Luttinger1946} (LT), who introduced a theory to determine the ground state of translationally invariant systems. While the construction scheme provided by LT can be extended beyond dipolar-coupled systems~\cite{Kaplan2007}, its original purpose was to find the ground-state configuration of arrangements of classical dipoles on lattices such as those in the paramagnetic salts. The ground-state configuration was found to be strongly dependent on the geometry of the lattice, and it is even sample-shape dependent for ferromagnetic alignment of the spins~\cite{Luttinger1946,Griffiths1964,Politi2006}. While the LT construction scheme does not apply to all lattices, it enables the determination of the ground-state configuration of common systems such as dipolar-coupled spins placed on the square lattice~\cite{Belobrov1983}. 

Remarkably, dipolar-coupled spin systems exhibit a continuous ground-state degeneracy in many different geometries~\cite{Belobrov1983,Prakash1990,Rastelli2002b,Way2018,Schonke2015}. The origin of this degeneracy is still not fully understood, although it has become clear that the degeneracy is not protected by symmetry so that even small perturbations, such as temperature or disorder, lift the degeneracy entirely through an \emph{order-by-disorder} transition~\cite{Henley1989a,Prakash1990}.

In recent years, the interest in dipolar systems has increased due to experimental work on the pyrochlore spin ices~\cite{Melko2003,Yavorskii2008}, leading to theoretical studies on systems with similar spin arrangements~\cite{Maksymenko2015,Holden2015,Way2018}. Furthermore, the desire to better understand the physics governing the spin-ices provided the motivation to explore correlated magnetic behavior in artificial spin systems with nanomagnetic moments taking on the role of the spins~\cite{Heyderman2013,Nisoli2013,Marrows2016}. Such artificial spin systems are, in contrast to the pyrochlores, neither restricted in lattice geometry nor the single particle magnetic anisotropy. Therefore, even though for initial investigations the focus was on Ising degrees of freedom~\cite{Wang2006,Farhan2013,Anghinolfi2015a,Sendetskyi2016}, there has since been an increased interest in nanomagnets with continuous degrees of freedom~\cite{Ostman2017,Leo2018,Streubel2018}.

\begin{figure}
    \centering
    \tikzstyle{decision} = [diamond, draw, fill=      ColorDecision,  text width=4.5em, text centered, inner sep=0pt, font=\scriptsize]
\tikzstyle{known} = [rectangle, draw, fill=       ColorKnown,     text width=8em, text centered, rounded corners, minimum height=4em, font=\scriptsize]
\tikzstyle{new}   = [rectangle, draw, fill=       ColorNew,       text width=4.5em, text centered, rounded corners, minimum height=4em, font=\scriptsize]
\tikzstyle{negconsequence} = [rectangle, draw, fill= ColorConsequence, text centered, rounded corners, minimum height=4em, font=\scriptsize]
\tikzstyle{consequence} = [rectangle, draw, fill= ColorNew, text centered, rounded corners, minimum height=4em, font=\scriptsize]
\tikzstyle{line} = [draw, -latex']

\begin{tikzpicture}[node distance = 2.5cm, auto]
    \node [known]                          (init)           {Classical dipolar-coupled spin system};
    \node [decision, below of=init]        (Q-LT)           {Luttinger-Tisza method~[8,9] works?};
    \node [decision, below of=Q-LT, node distance=3.0cm]        (Q-comm)         {Commen\-surate order?};
    \node [decision, below of=Q-comm]      (QEr)            {Is \vecrep irreducible?};
    \node [decision, right of=QEr, node distance=3.0cm]         (Q1Do)           {Decays to only 1D irreps?};
    \node [decision, below of=Q1Do, node distance=3.0cm]        (n1Do)           {Has lowest energy block $\dblock>1$?};
    \node [new,   below of=n1Do]           (contDegdbar)    {Continuous degeneracy of dimension $\dblock$};
    \node [new,   left of=contDegdbar,node distance=3.0cm]            (yEr)            {Continuous degeneracy of dimension $d$};
    \node [new, right of=contDegdbar]    (nCD)            {No continuous degeneracy};
    \node [negconsequence,text width=2cm, right of=Q-comm, node distance=3.0cm] (complicated)    {No general\\ statement possible};
    \path [line] (init) -- (Q-LT);
    \path [line] (Q-LT) -| node [very near start] {no} (complicated);
    \path [line] (Q-LT) -- node {yes} (Q-comm);
    \path [line] (Q-comm) -- node {yes} (QEr);
    \path [line] (Q-comm) -- node [near start] {no} (complicated);
    \path [line] (QEr) -- node [near start] {yes} (yEr);
    \path [line] (QEr) -- node [near start] {no} (Q1Do);
    \path [line] (Q1Do) -| node [near start] {yes} (nCD);
    \path [line] (Q1Do) -- node [near start] {no} (n1Do);
    \path [line] (n1Do) -- node [near start] {yes} (contDegdbar);
    \path [line] (n1Do) -| node [near start] {no} (nCD);
   
\end{tikzpicture}
    \caption{(color online) This flow diagram summarizes the findings of this article. Applying this scheme to a generic classical dipolar-coupled spin system, the LT method is extended with an additional classification based on reduction of the vector representation \vecrep in the point symmetry group \pointsym, which determines the nature of the ground-state degeneracy.}
    \label{fig:flow-diagram}
\end{figure}
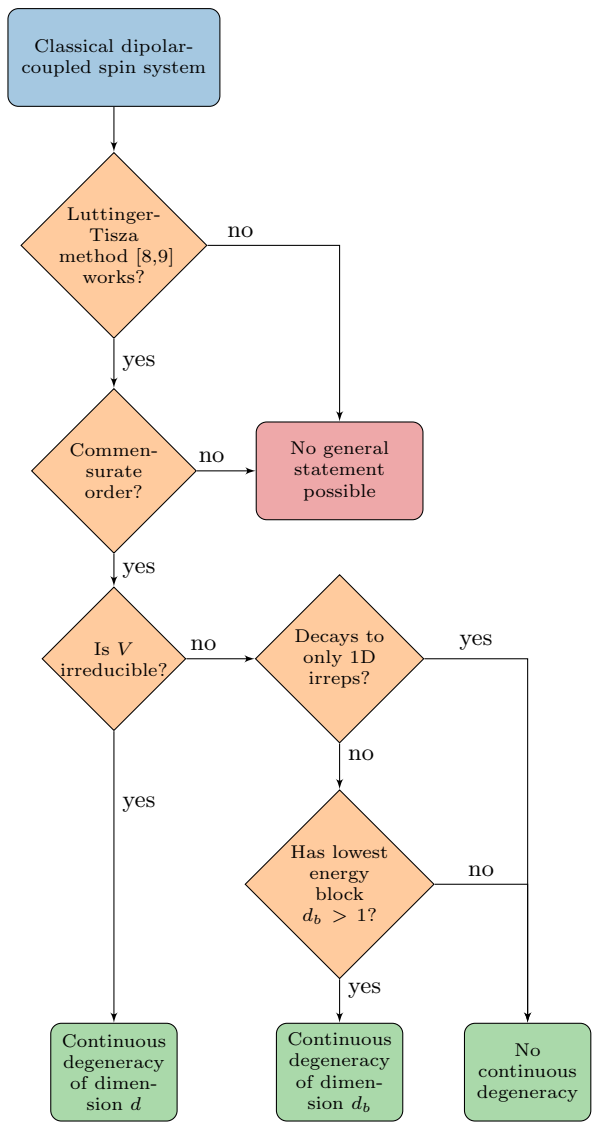

The theory for the artificial spin systems with continuous degrees of freedom, experimentally addressed in in Refs.~\cite{Leo2018,Streubel2018}, has been discussed in previous works~\cite{Belobrov1983,Prakash1990}. However, the field lacks a generalization that is free from assuming a specific lattice. In this paper, we provide a more general approach via a detailed symmetry discussion, which gives a framework to determine the ground-state degeneracy for some generic lattices. This leads to a guide for the determination of whether a particular classical dipolar system has a continuous ground-state degeneracy. We provide the essence of this discussion in the flow diagram shown in Fig.~\ref{fig:flow-diagram}.

The remainder of this paper is structured as follows: In Section~\ref{sec:model}, the model of classical dipolar spins is introduced through the Hamiltonian with an emphasis on symmetries. After a brief review of the LT method in Section~\ref{sec:lt}, we extend this method in Section~\ref{sec:pointsym} using the representation theory for the point symmetry group of the lattice to determine the ground-state degeneracy. We then illustrate this method for several examples in Section~\ref{sec:examples}. Finally, we summarize our results in Section~\ref{sec:conclusion}, where we give an outlook on how the method presented here can be generalized to include the order-by-disorder transitions commonly found in dipolar-coupled systems. 

\section{Model \& Symmetries}
\label{sec:model}
Here, we introduce the classical model of dipolar-coupled spins with the Hamiltonian
{
\begin{align} 
        H&= \frac{D}{2}\sum_{i\neq j}\frac{1}{|\vec r_{ij}|^3}\left[ \vec S_i\cdot \vec S_j - 3 \left( \vec S_i\cdot \hat r_{ij} \right)\left( \vec S_j\cdot \hat r_{ij} \right) \right], 
    \label{eq:hamiltonian} 
\end{align} 
}%
where $D$ is the dipolar interaction strength, defined for $|\vec S_i|=1$. The vector $\vec r_{ij}$ is the difference vector between the positions of the sites $i$ and $j$ on a regular lattice, and $\hat r_{ij}$ is the normalization of $\vec r_{ij}$ to unit length.

A classical spin $\vec S_i$ is typically described by a vector on the unit sphere, i.e., a Heisenberg spin. However, additional anisotropies can lower the effective degree of freedom of the spins. For example, in artificial spin ice, shape anisotropy can give rise to Ising-like behavior~\cite{Wang2006} or XY-like behavior~\cite{Leo2018,Streubel2018}, and the presence of magnetocrystalline anisotropy can result in clock-model-like behavior~\cite{Louis2018}. For the remainder of the article, we will focus on spins with XY or Heisenberg behavior, in order to determine when continuous ground-state degeneracies arise.

Regardless of whether the spin is Heisenberg or XY, the dipolar Hamiltonian~\eqref{eq:hamiltonian} is geometrically frustrated. Namely, the first term $ \vec S_i\cdot \vec S_j$ is minimized for antiparallel spin alignment, whereas the second term $-3( \vec S_i\cdot \hat r_{ij})( \vec S_j\cdot \hat r_{ij} )$ is minimized for parallel alignment if the spins can align along their bond. As a consequence, in a system with dipolar interactions given by Eq.~\eqref{eq:hamiltonian}, the alignment of spins follows the ``head-to-tail'' rule, i.e., spins align parallel if they can align along their bond, and antiparallel if the spins are orthogonal to the bond. 

In dimension, $d=3$, the $r^{-3}$ dipolar interaction is long-range and, if the local magnetic configuration has a net magnetic moment, as in a ferromagnet, the energy density will grow with system size. As a result, sample-shape dependent corrections are expected~\cite{Mukamel2009} and Weiss domains are formed through the minimization of the stray field energy. This sensitivity to the sample shape of ``ferromagnetic'' configurations of dipolar systems is a result of Griffiths' theorem~\cite{Griffiths1964}. This theorem implies that ``ferromagnetic'' configurations cannot be single domain in the thermodynamic limit since sample-shape dependent corrections in the form of the demagnetization factor arise. However, if the local magnetic configuration does not have a net magnetic moment, no demagnetization factor arises and the magnetic stray fields typically self-screen such as in an antiferromagnet or in the dipolar spin ice model~\cite{Melko2003}. For the remainder of this paper, we neglect the influence of sample-shape dependent corrections and boundary terms as in previous studies~\cite{Belobrov1983,Prakash1990}. Thus, for $d=3$, the generality of our work is restricted to ``non-ferromagnetic'' systems. However, it is noted that even low-dimensional $d<3$ dipolar-coupled systems with a ferromagnetic ground state, such as XY spins on the triangular lattice~\cite{Malozovsky1991}, are known to be sensitive to the sample shape or the truncation of the Hamiltonian~\cite{Politi2006}.

In addition to the long-range nature of the dipolar Hamiltonian in Eq.~\eqref{eq:hamiltonian}, the Hamiltonian also possesses the symmetry group $\Z_2 \times \transsym\times \pointsym$, which is rather unusual for spin Hamiltonians. In this group, the time-reversal symmetry is reflected by $\Z_2$, the translational invariance by \transsym, and \pointsym corresponds to the point-group symmetry of the lattice. The time-reversal symmetry follows directly from the invariance of Eq.~\eqref{eq:hamiltonian} under $\vec S_i\mapsto -\vec S_i$. The translational invariance \transsym is explicitly given by the mapping
\begin{align}
    (\vec r_i, \vec S_i)&\stackrel{\transsym}{\mapsto } (\vec r_{i'},
    \vec S_{i'})=(\vec{r_i}-\vec{t},\vec{S}_{i'}).
\label{eq:translational-symmetry-action}
\end{align}
Since only relative coordinates appear in the dipolar Hamiltonian in Eq.~\eqref{eq:hamiltonian}, a shift of the system by a vector $\vec{t}$ is irrelevant whenever $\vec t$ is a lattice vector.
%
Finally, the point symmetry group \pointsym is inherited by the underlying lattice.  If we denote the vector representation of \pointsym in the $d$-dimensional vector space with \vecrep, where $d$ is the dimension of a spin, then a vector $\vec v\in\mathbb R^d$ transforms under the action of $g\in\pointsym$ according to $\vecrep(g)\vec v$, such that the Hamiltonian~\eqref{eq:hamiltonian} stays invariant under
\begin{align}
    (\vec r_i,\vec S_i)&\stackrel{\pointsym}{\mapsto} (\vecrep(g) \vec r_i,
    \vecrep(g) \vec S_i).
    \label{eq:point-symmetry-action}
\end{align}
Here, \vecrep acts on both the lattice and the spin simultaneously. This simultaneous action of \pointsym on both vectors $\vec r_i $ and $\vec S_i$ is required by the second term in the Hamiltonian given in Eq.~\eqref{eq:hamiltonian}, which tightly connects real-space and spin-space. We discuss specific examples of complete symmetry groups in Section~\ref{sec:examples}.

Formally, the model incorporates two different dimensions, the real-space lattice dimension $d_{\text{lattice}}$ and the spin-space dimension $d_{\text{spin}}$. The two spaces are coupled as a result of the dipolar interaction described by the Hamiltonian given by Eq.~\eqref{eq:hamiltonian}. Hence, it is useful to introduce a working dimension $d$, which is the dimension of the space in which both the spins and the lattice can be embedded. For some simple situations, it can be sufficient to work in the smaller of the two spaces. This can be seen, for example, for in-plane XY spins on the cubic lattice. Here, the XY anisotropy reduces the point symmetry group of the system to the point symmetry group of the square lattice. Therefore, the problem of XY spins on a cubic lattice reduces to the problem of XY spins on square-lattice layers~\cite{Belobrov1983}.

\section{Ground States}
\label{sec:gs}
In this section, we use the symmetries of the dipolar Hamiltonian to explain the origin of the ground-state degeneracy. For this purpose, we first summarize the LT method~\cite{Luttinger1946} and subsequently extend the LT method by using the representation theory for the point symmetry group to determine the nature of the ground-state degeneracy.

\subsection{Luttinger-Tisza construction}
\label{sec:lt}
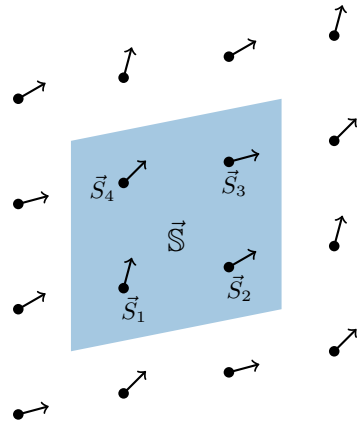
\begin{figure}
    \centering
    \newcommand{\spin}[3]{
    \draw[->,thick, rotate around={#3:(#1,{#2+.2*(#1)})}] (#1,{#2+.2*(#1)}) -- ($(#1,{#2+.2*(#1)})+(.3,0)$);
    \fill (#1,{#2+.2*(#1)}) circle (.05);
}
\newcommand{\ltFigureAnglea}{30}
\newcommand{\ltFigureAngles}{45}
\newcommand{\ltFigureAngled}{75}
\newcommand{\ltFigureAnglef}{15}
\begin{tikzpicture}[scale=1.4]
    \fill[color = ColorKnown] (-.5,-.6) -- (1.5,-.2) -- (1.5,1.8) -- (-.5,1.4) -- cycle;
    \foreach \x in {-1,1}
    \foreach \y in {-1,1}
    {
        \spin{\x  }{\y+1}{\ltFigureAnglea}
        \spin{\x+1}{\y  }{\ltFigureAngles}
        \spin{\x+1}{\y+1}{\ltFigureAngled}
        \spin{\x  }{\y  }{\ltFigureAnglef}
    }
    \node at (0.1,-.20) {$\vec S_1$};
    \node at (1.1,0) {$\vec S_2$};
    \node at (1.05,1.) {$\vec S_3$};
    \node at (-.2,.95) {$\vec S_4$};
    \node at (.5,.5) {\large $\S$};
\end{tikzpicture}
    \caption{(color online) Schematic illustrating the LT method, which is based on an ansatz for the magnetic unit cell (shaded in blue). The magnetic unit cell contains the spins $\vec S_1,\dots,\vec S_N$ (here $N=4$), which form the effective spin configuration $\S$. The aim of the LT method is to provide the so-called basic arrays, namely a symmetry-guided basis for \S.}
    \label{fig:lt-explanatory-figure}
\end{figure}

The LT method is based on an ansatz for the magnetic unit cell that stays invariant under lattice symmetry operations (\transsym and \pointsym) and subsequently minimizes the dipolar energy associated with the magnetic unit cell. The generalized LT method builds on the original method and is based on a Fourier transformation of the interaction, and finding the ordering vector of the ground state in the Fourier space rather than in real-space~\cite{Luttinger1946,Kaplan2007}. In general, commensurate and incommensurate order can be found using the generalized method, although the original LT method is only applicable to systems with commensurate order, irrespective of the unit cell size. Indeed, the generalized LT method finds commensurate order for many dipolar systems such as those listed in Table~\ref{tab:overview-systems}. For these systems, the magnetic unit cell is at most double the structural unit cell.

If the LT method can successfully be applied to a dipolar-coupled spin system, then minimization of the dipolar energy for a suitable magnetic unit cell leads to the exact ground-state of the system. When unphysical solutions appear, then the LT method fails. We will discuss the issue of unphysical solutions towards the end of this section after introducing the use of the LT method for finding the ground-state of dipolar-coupled spin systems.

For a general dipolar-coupled spin system, one starts by making an ansatz for the magnetic unit cell that respects the point symmetry group of the lattice. Subsequently, the $N$ spins in the magnetic unit cell are collected into one vector $\S=(\vec S_1,\vec S_2,\dots,\vec S_N)$ as illustrated in Fig.~\ref{fig:lt-explanatory-figure}. Since the Hamiltonian in Eq.~\eqref{eq:hamiltonian} is quadratic, the effective Hamiltonian for the magnetic unit cell can be written in terms of \S as $H=-\S^\dagger \mathcal H \S$, where $\mathcal H\S$ is the induced dipolar field of the configuration \S. This finite-dimensional diagonalization problem is further simplified by taking into account the translational invariance \transsym of the Hamiltonian: Using the representation theory for the translational invariance in the magnetic unit cell, one can obtain a symmetry-guided basis for \S, the so-called basic arrays. This basis is constructed using the irreducible representations of \transsym in the magnetic unit cell, which correspond to the discrete Fourier states. Therefore, typical basic arrays are, for example, the ferromagnetic configuration 
($\S_{ \ferro ,x}=(\hat e_x, \hat e_x,\dots)$) or the antiferromagnetic configuration ($\S_{\afm,x}=(\hat e_x, -\hat e_x,\dots)$)~\footnote{In the paper of Luttinger and Tisza, the configuration $\S_{\ferro,z}=Z_1$ and $\S_{\afm,z}=Z_8$, respectively}.
These symmetry-guided configurations are mutually orthogonal by construction and, because of the translational invariance, the different sectors such as the ferromagnetic sector (\ferro) or the antiferromagnetic sector (\afm) are not mixed. This leads to a further simplification of the ground state, since $\mathcal H=\oplus_i \mathcal H_i$, i.e., $\mathcal H$ is block-diagonal. Each of the blocks $\mathcal H_i$ describes the coupling between the basic arrays with one type of ordering. For example, one block describes the coupling between the ferromagnetic configurations $\S_{\ferro,x},\S_{\ferro,y},\dots$ and another describes the coupling between antiferromagnetic configurations $\S_{\afm,x},\S_{\afm,y},\dots$. Hence, each of the blocks is $d$-dimensional. Therefore, with the LT method, we find
\begin{align}
    \mathcal H&=
            \begin{pmatrix}
                \blockmatrix{${\mathcal H}_{\ferro}$}& 0& \dots\\
                0&\blockmatrix{${\mathcal H}_{\afm}$} & \dots\\
                \vdots& \vdots& \ddots\\
            \end{pmatrix},
    \label{eq:block-diagonal-natural}
\end{align}
which significantly simplifies the problem since only a small number of explicit lattice summations have to be carried out. Here, it should be noted that, if the magnetic unit cell size is increased, then $\mathcal H$ becomes larger, but this method can still be applied.

Finally, the LT method only guarantees the ``weak condition'' $\S^2=\sum_i^N\vec S_i^2= N$, and can therefore give unphysical solutions where the ``strong condition'' $|\vec S_i|=1$ is violated.  If an unphysical lowest-energy configuration \S is identified by this method, then the method fails to provide the ground state. For such systems, one can either introduce Lagrange-multipliers~\cite{Litvin1974,Kaplan2007} or resort to numerical methods~\cite{Way2018,Maksymenko2015,Holden2015}. While Lagrange-multipliers render the problem non-linear, using numerical methods one typically finds non-orthogonal states as ground-state configurations. When the method fails, it is not clear if the system possesses a continuous degeneracy~\cite{Way2018} or a discrete degeneracy~\cite{Holden2015,Maksymenko2015}. Nevertheless, as seen from Table~\ref{tab:overview-systems}, the LT method works for many important systems, and we show in the next section that, for these systems, \pointsym uniquely defines the type and dimension of the degeneracy.

\begin{table}
    \centering
    \caption{Overview of previous theoretical treatments of dipolar-coupled coupled spins placed at the sites of various lattices and whether the ground state can be determined by the LT method.}
\begin{tabular}{l|r}
    \hline\hline
        Lattice&LT?\\\hline
             chain lattice         ~\cite{Belobrov1983}              & Yes \\
             rectangular lattice   ~\cite{Belobrov1983}              & Yes \\
             square lattice        ~\cite{Belobrov1983,Prakash1990}  & Yes \\
             honeycomb lattice     ~\cite{Prakash1990}               & Yes \\
             kagome lattice        ~\cite{Holden2015,Maksymenko2015} & No  \\
             cubic lattice         ~\cite{Luttinger1946,Belobrov1983}& Yes \\
             ``fcc-kagome'' lattice~\cite{Way2018}                   & No  \\
             \hline\hline
        \end{tabular}
    \label{tab:overview-systems}
\end{table}

\subsection{Continuous ground-state degeneracy}
\label{sec:pointsym}
The point symmetry group \pointsym determines the type of degeneracy in the following way: Since \pointsym is a symmetry of the Hamiltonian, as described by Eq.~\eqref{eq:point-symmetry-action}, it is therefore also a symmetry of the effective interaction matrix $\mathcal H$. Hence, symmetry-group operations have to commute with $\mathcal H$, formally expressed as $[R(g),\mathcal H]=R(g) \mathcal H- \mathcal HR(g)=0$ for all point symmetry group elements $g\in \pointsym$, where $R$ is a representation of \pointsym. The representation $R$ can be found considering that each block matrix $\mathcal H_i\in\{\mathcal H_\ferro, \mathcal H_\afm,\dots\}$ has dimension $d$. Indeed, given one spin, for example $\vec S_1$, all other spins in the magnetic unit cell are defined by the index $i\in \{\ferro,\afm,\dots\}$. Therefore the representation of \pointsym acting on the subspace for $\mathcal H_i$ is \vecrep, the vector representation of \pointsym. Hence, the representation for the entire matrix $\mathcal H$ is given by $R=\oplus_N \vecrep$.

The symmetry condition implies that, for one block matrix, the reduced symmetry condition is $[\vecrep(g),\mathcal H_i]=0$ for all $g\in \pointsym$. For the case where \vecrep is irreducible, the first lemma of Schur implies that $H_i=h_i\identity$ so that there are $d$ mutually orthogonal configurations $\S_1,\dots,\S_d$, all having the same energy. Hence, any superposition $\S_{\vec \alpha}=\sum_{i=1}^d \alpha_i \S_i$, with the normalization constraint $\sum_{i=1}^d |\alpha_i|^2 =1$, yields the same energy as the basis states since the Hamiltonian from Eq.~\eqref{eq:hamiltonian} is quadratic in $\vec S_i$. The normalization constraint itself is the equation of a $(d-1)$-dimensional sphere. Hence, the ground-state manifold is described by a $(d-1)$-dimensional sphere.

For the case where \vecrep is reducible, the block matrices $\mathcal H_i$ decompose into smaller block matrices. The explicit summation over the lattice identifies the smaller block matrix with dimension \dblock{} that is lowest in energy. Then the ground-state manifold is described by the reduced $(\dblock-1)$-dimensional sphere. If $\dblock=1$, the degeneracy is described by the $0$-sphere, which is equivalent to $\Z_2$ and therefore only a discrete degeneracy is recovered and not a continuous degeneracy that is found for systems where $\dblock>1$.

To conclude, the use of the representation theory, not only for the translational invariance but also for the point symmetry group, leads to a more generic treatment than that implemented by Luttinger and Tisza. Even though Luttinger and Tisza used the point symmetry group to simplify their problem, their approach did not exploit the representation theory for the point symmetry group. In contrast, the extension presented here uses the representation theory for both the translational invariance and the point symmetry group, so that continuous ground-state degeneracies appear naturally. Furthermore, the continuous degeneracy is not accidental, as it does not require a fine-tuning of parameters, but instead follows from symmetry. Thus, the degeneracy is not guaranteed by a continuous symmetry of the Hamiltonian, but rather follows from the finite point symmetry group.

\section{Examples}
\label{sec:examples}
We now illustrate the concepts presented in Section~\ref{sec:gs} with some examples. Specifically, we consider the dipolar-coupled XY spins on the square lattice in Section~\ref{sec:ex:square}, Heisenberg spins on the (tetragonally distorted) cubic lattice in Section~\ref{sec:ex:cubic}, and XY spins on the triangular lattice in Section~\ref{sec:ex:triangular}.

\subsection{XY spins on the square lattice}
\label{sec:ex:square}
Here we determine the ground-state of dipolar-coupled XY spins on the square lattice, as this example has already been well studied~\cite{Belobrov1983,Prakash1990,DeBell1997,Baek2011,Fernandez2007,Carbognani2000,Patchedjiev2005,LeBlanc2006,DeBell1997,Schildknecht2018a,Leo2018,Streubel2018}. Here, it is expected that the ground state exhibits a continuous degeneracy equivalent to the $1$-sphere, independent of whether a truncation is applied to the Hamiltonian~\cite{Prakash1990} or not~\cite{Belobrov1983}. The symmetry group of this system is given by $\Z_2\times \transsym_{sq}\times C_{4v}$, where $\Z_2$ is the time-reversal symmetry and $C_{4v}$ is the point symmetry group of the square lattice. The translational invariance $\transsym_{sq}$ can be parameterized via vectors $\vec t=x\hat e_x+y\hat e_y$, with $x,y\in \Z$ and $\hat e_x, \hat e_y$ being the unit vectors along the $x$-axis and the $y$-axis, respectively. Therefore, $\transsym_{sq}$ is isomorphic to $\Z\times\Z$.

\begin{figure}
    \subfloat[][$(0,0)$]{
        \begin{tikzpicture}
    \draw[-,color=white] (-.5,-.5) -- (-.5, 1.5) -- ( 1.5, 1.5) -- (1.5,-.5);
    \draw[->,thick] ($(0,0)+(0,-.3)$) -- ($(0,0)+(0,.3)$);
    \draw[->,thick] ($(0,1)+(0,-.3)$) -- ($(0,1)+(0,.3)$);
    \draw[->,thick] ($(1,0)+(0,-.3)$) -- ($(1,0)+(0,.3)$);
    \draw[->,thick] ($(1,1)+(0,-.3)$) -- ($(1,1)+(0,.3)$);
\end{tikzpicture}
        \label{fig:square-basic-arrays1}
    }\hskip.1\textwidth
    \subfloat[][$(0,\pi)$]{
        \begin{tikzpicture}
    \draw[-,color=white] (-.5,-.5) -- (-.5, 1.5) -- ( 1.5, 1.5) -- (1.5,-.5);
    \draw[->,thick] ($(0,0)+(0,-.3)$) -- ($(0,0)+(0,.3)$);
    \draw[<-,thick] ($(0,1)+(0,-.3)$) -- ($(0,1)+(0,.3)$);
    \draw[->,thick] ($(1,0)+(0,-.3)$) -- ($(1,0)+(0,.3)$);
    \draw[<-,thick] ($(1,1)+(0,-.3)$) -- ($(1,1)+(0,.3)$);
\end{tikzpicture}
        \label{fig:square-basic-arrays2}
    }\hskip.1\textwidth
    \subfloat[][$(\pi,0)$]{
        \begin{tikzpicture}
    \draw[-,color=white] (-.5,-.5) -- (-.5, 1.5) -- ( 1.5, 1.5) -- (1.5,-.5);
    \draw[->,thick] ($(0,0)+(0,-.3)$) -- ($(0,0)+(0,.3)$);
    \draw[->,thick] ($(0,1)+(0,-.3)$) -- ($(0,1)+(0,.3)$);
    \draw[<-,thick] ($(1,0)+(0,-.3)$) -- ($(1,0)+(0,.3)$);
    \draw[<-,thick] ($(1,1)+(0,-.3)$) -- ($(1,1)+(0,.3)$);
\end{tikzpicture}
        \label{fig:square-basic-arrays3}
    }\hskip.1\textwidth
    \subfloat[][$(\pi,\pi)$]{
        \begin{tikzpicture}
    \draw[-,color=white] (-.5,-.5) -- (-.5, 1.5) -- ( 1.5, 1.5) -- (1.5,-.5);
    \draw[->,thick] ($(0,0)+(0,-.3)$) -- ($(0,0)+(0,.3)$);
    \draw[<-,thick] ($(0,1)+(0,-.3)$) -- ($(0,1)+(0,.3)$);
    \draw[<-,thick] ($(1,0)+(0,-.3)$) -- ($(1,0)+(0,.3)$);
    \draw[->,thick] ($(1,1)+(0,-.3)$) -- ($(1,1)+(0,.3)$);
\end{tikzpicture}
        \label{fig:square-basic-arrays4}
    }
    \caption{The four basic arrays for the two-by-two magnetic unit cell on the square lattice with moments aligned along the $y$-axis. The Fourier vector that generates the basic array is indicated below each figure. The lattice summations associated with the dipolar Hamiltonian reveal that the configuration shown in (c) has the lowest energy.}
    \label{fig:square-basic-arrays}
\end{figure}
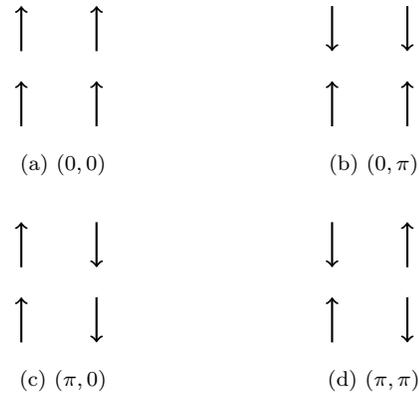

In the next step, the LT method is applied to a two-by-two magnetic unit cell, so that $\S=(\vec S_1, \vec S_2,\vec S_3,\vec S_4)$. Since $C_4$ is a symmetry of the system, it is sufficient to only consider basic arrays with spins parallel or antiparallel to $\hat e_y$. The LT method then suggests a suitable basis based on the translational invariance $\transsym_{sq}$, which is, however, broken by the two-by-two magnetic unit cell. Hence, the basic arrays correspond to (discrete) Fourier components that arise due to the reduced translational invariance. Since the translational invariance is reduced by a factor of two in every direction, the basic arrays are formed by the square root of unity in every direction. The resulting basic arrays are depicted in Fig.~\ref{fig:square-basic-arrays}, with the Fourier vector that characterizes the elements given below each figure. From explicit calculation, it can be observed that the configuration depicted in Fig.~\ref{fig:square-basic-arrays3} is the basic array with the lowest energy~\cite{Belobrov1983}.

\begin{figure}
    \centering
    \includegraphics[width=.6\columnwidth]{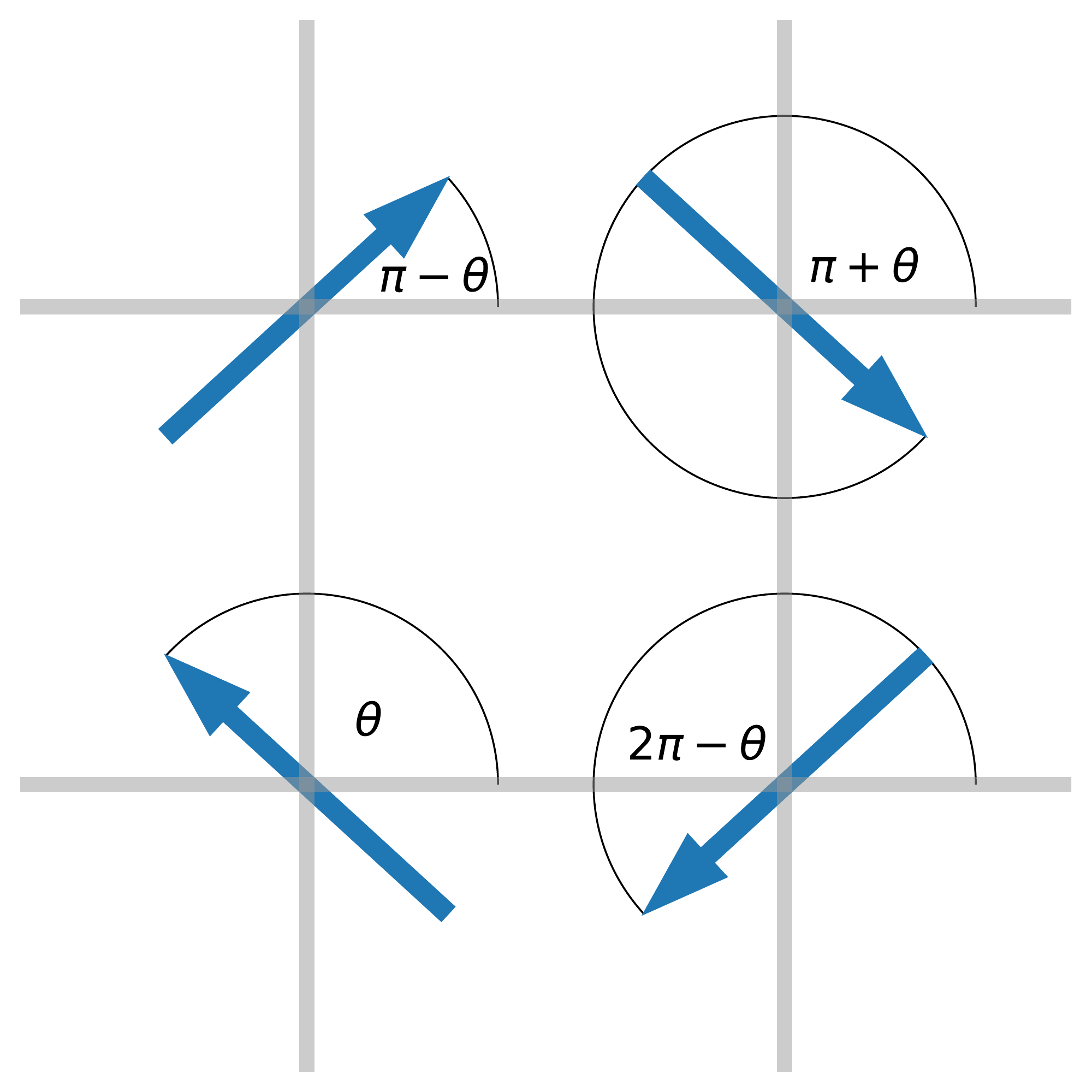}
    \caption{(color online) The magnetic unit cell for the ground state of dipolar-coupled XY
    spins on the square lattice for a general $\theta$.}
    \label{fig:gs-square}
\end{figure}

Finally, we need to validate if \vecrep is irreducible (for details of how this is done, see for example Ref.~\cite{Tinkham1964}). The reduction is shown in Table~\ref{table:vecrep_reduction}, and we indeed observe that $\vecrep\equiv E$ is irreducible in $C_{4v}$. Hence, we know that the basic arrays corresponding to Fig.~\ref{fig:square-basic-arrays3}, with spins either aligned along $\hat e_x$ or aligned along $\hat e_y$, have the same energy. Hence, we have found a continuous ground-state degeneracy described by the $1$-sphere, which is depicted in Fig.~\ref{fig:gs-square}, in agreement with previous studies~\cite{Belobrov1983,Prakash1990}.

\begin{table}
    \centering
    \caption{Character table for the point symmetry group of the square lattice $C_{4v}$ and the reduction of \vecrep in this group.}
    \begin{tabular}{c|rrrrr|c}
        \hline\hline
        $C_{4v}$& $\identity$& $2C_4$& $C_2$& $2\sigma_h$& $2\sigma_{d}$\\\hline
        $A_1$   & $1$ & $1$  & $1$  & $1$  & $1$\\
        $A_2$   & $1$ & $1$  & $1$  & $-1$ & $-1$\\
        $B_1$   & $1$ & $-1$ & $1$  & $1$  & $-1$\\
        $B_2$   & $1$ & $-1$ & $1$  & $-1$ & $1$\\
        $E$     & $2$ & $0$  & $-2$ & $0$  & $0$    & \\\hline
        \vecrep & $2$ & $0$  & $-2$ & $0$  & $0$    & $\equiv E$\\
        \hline\hline
    \end{tabular}
    \label{table:vecrep_reduction}
\end{table}

\subsection{Heisenberg spins on the (distorted) cubic lattice}
\label{sec:ex:cubic}
\begin{figure}
    \centering
    \includegraphics[width=.6\columnwidth]{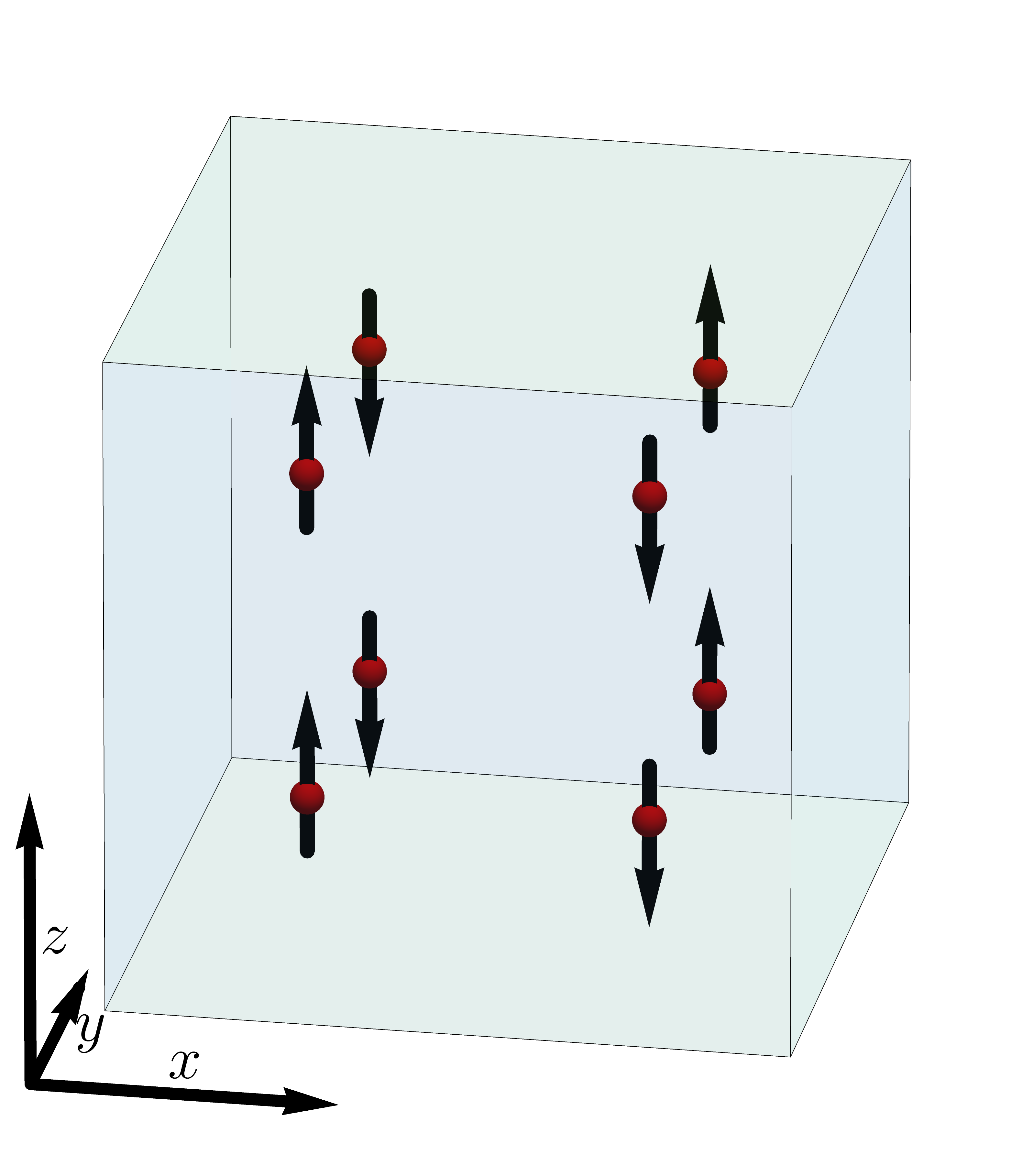}
    \caption{(color online) Ground-state configuration for dipolar-coupled Heisenberg spins on the cubic lattice according to Ref.~\cite{Luttinger1946}. The shaded area indicates the magnetic unit cell.}
    \label{fig:cubic-striped}
\end{figure}

To provide a higher dimensional example, we consider dipolar-coupled Heisenberg spins on the cubic lattice with lattice constants $a=b=c$. Here, a continuous ground-state degeneracy described by the $2$-sphere is expected independent of whether a truncation is applied to the Hamiltonian~\cite{Romano1994b,Chamati2014} or not~\cite{Belobrov1983,Chamati2016}. The symmetry of this system is given by $\Z_2\times \transsym_{cu}\times O_h$ where $\Z_2$ is time reversal, $O_h$ is the point symmetry group of the cubic lattice and $\transsym_{cu}\cong \Z\times\Z\times\Z$ parametrizes the translational invariance with vectors $\vec t=x\hat e_x+y\hat e_y +z\hat e_z$. Subsequently, the LT method requires the evaluation of $2^3=8$ lattice summations analogous to the ones for the square lattice~\cite{Luttinger1946}. From this calculation, Luttinger and Tisza found that the striped configuration depicted in Fig.~\ref{fig:cubic-striped} has the lowest energy of all of the basic arrays. Finally, we reduce the vector representation \vecrep in the group $O_h$ in Table~\ref{tab:character-table-oh}, which results in $\vecrep\equiv T_{1u}$, so that the vector representation is once more irreducible. This yields a continuous ground-state degeneracy corresponding to the $2$-sphere in accordance with previous studies~\cite{Belobrov1983}. In Ref.~\cite{Belobrov1983} a graphical representation of this ground-state manifold is provided in their Fig.~1.

\begin{table}
    \centering
    \caption{ Character table for point group $O_h$  and the reduction of \vecrep in this group.}
    \begin{tabular}{c|rrrrrrrrrr|c}
        \hline\hline
$O_{h}$&
$\identity$&
$8C_{3}$&
$6C_{2}$&
$6C_{4}$&
$3C_{4}^2$&
$i$&
$6S_{4}$&
$8S_{6}$&
$3\sigma_{h}$&
$6\sigma_{d}$

\\
\hline
$A_{1g}$&
$1$&
$1$&
$1$&
$1$&
$1$&
$1$&
$1$&
$1$&
$1$&
$1$

\\
$A_{2g}$&
$1$&
$1$&
$-1$&
$-1$&
$1$&
$1$&
$-1$&
$1$&
$1$&
$-1$

\\
$E_{g}$&
$2$&
$-1$&
$0$&
$0$&
$2$&
$2$&
$0$&
$-1$&
$2$&
$0$

\\
$T_{1g}$&
$3$&
$0$&
$-1$&
$1$&
$-1$&
$3$&
$1$&
$0$&
$-1$&
$-1$

\\
$T_{2g}$&
$3$&
$0$&
$1$&
$-1$&
$-1$&
$3$&
$-1$&
$0$&
$-1$&
$1$

\\
$A_{1u}$&
$1$&
$1$&
$1$&
$1$&
$1$&
$-1$&
$-1$&
$-1$&
$-1$&
$-1$

\\
$A_{2u}$&
$1$&
$1$&
$-1$&
$-1$&
$1$&
$-1$&
$1$&
$-1$&
$-1$&
$1$

\\
$E_{u}$&
$2$&
$-1$&
$0$&
$0$&
$2$&
$-2$&
$0$&
$1$&
$-2$&
$0$

\\
$T_{1u}$&
$3$&
$0$&
$-1$&
$1$&
$-1$&
$-3$&
$-1$&
$0$&
$1$&
$1$

\\
$T_{2u}$&
$3$&
$0$&
$1$&
$-1$&
$-1$&
$-3$&
$1$&
$0$&
$1$&
$-1$
\\\hline
\vecrep&
$3$&
$0$&
$-1$&
$1$&
$-1$&
$-3$&
$-1$&
$0$&
$1$&
$1$&$\equiv T_{1u}$\\
\hline\hline
    \end{tabular}
    \label{tab:character-table-oh}
\end{table}

If the cubic lattice has a tetragonal distortion, where one lattice constant (e.g.\ $c$) is different from the other two ($a=b$), then the three-dimensional block matrix describing the ground state of the undistorted cubic lattice $\mathcal H_{\text{striped}}$ reduces to two block matrices of dimensions $1$ and $2$, respectively. To determine which block matrix is lower in energy, one can consider the two cases $c<a=b$ and $c>a=b$. If $c>a$, the system behaves as weakly interacting layers, that follow the symmetry constraints given by a square lattice. As a consequence, the two-dimensional representation is lower in energy, which yields a continuously degenerate ground state whose manifold resembles the unit circle. This is analogous to the manifold found for XY spins on a square lattice. If $c<a$, the system consists of chains of spins with weak interaction between the chains, whose low-energy sector is described by the one-dimensional block. Therefore, no continuous degeneracy emerges. Both of these cases are in agreement with previous literature~\cite{Belobrov1983}.

\subsection{XY spins on the triangular lattice} 
\label{sec:ex:triangular}
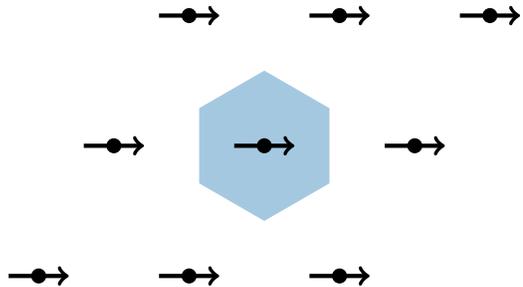
\begin{figure}
    \newcommand{\spin}[2]{
    \coordinate (a) at (1*#2+0.5*#1, 0*#2+0.8660254037844386*#1);
    \draw[->,ultra thick] 
    ($ (a) - (0.4,0) $)
    --
    ($ (a) + (0.4,0) $);
    \fill (a) circle (.1);
}
\begin{tikzpicture}
    \fill[color=ColorKnown] (-30:1) \foreach \x in {30,90,...,330} {  -- (\x:1) };

    \foreach \x in {-2,0,...,2}
    \foreach \y in {-2,0,...,2}
    {
        \spin{\x}{\y}
    }
\end{tikzpicture}
    \caption{(color online) Ground-state configuration for dipolar-coupled XY spins on the triangular lattice according to Ref.~\cite{Malozovsky1991}. The shaded area indicates the magnetic unit cell, which is also the structural unit cell due to the ferromagnetic ground-state configuration. }
    \label{fig:triangular}
\end{figure}

As a third example, dipolar-coupled XY spins on the triangular lattice are considered. If no truncation is applied to the Hamiltonian, the system orders ferromagnetically~\cite{Malozovsky1991}. Therefore, the configuration depicted in Fig.~\ref{fig:triangular} is one of the ground states of the system. As the local ground state is ferromagnetic, a finite sample will break up into domains in the ground state. As in previous literature~\cite{Malozovsky1991,Rastelli2002b,Tomita2009}, however, the effect of the sample shape is neglected here, and we only consider the local configuration consisting of a single domain.

Furthermore, in contrast to the previous examples given in this section, only the non-truncated Hamiltonian can be considered, since otherwise the ground-state configuration is altered~\cite{Politi2006,Tomita2009}. Indeed, when a truncation is applied to the Hamiltonian, the ground state contains a magnetic structural length scale that depends on the truncation, and it can no longer be derived by the LT method~\cite{Politi2006}. Therefore, the discussion in this section is restricted to the case where no truncation is applied to the Hamiltonian.

In this case, the ground-state configuration of dipolar-coupled XY spins on the triangular lattice is ferromagnetic~\cite{Malozovsky1991}. This means that the structural and the magnetic unit cell are the same so that the strong and the weak condition of the LT method are equivalent. Since the LT method works in this case and a commensurate ordering is obtained, the next step according to Fig.~\ref{fig:flow-diagram} is to determine if \vecrep is irreducible. Using the character table for the point symmetry group of the triangular lattice, $C_{6v}$ (see Table~\ref{tab:character-table-c6v}), one finds that $\vecrep \equiv E_1$ is irreducible. Hence, a continuous ground-state degeneracy described by a $1$-sphere is found, in agreement with previous studies~\cite{Rastelli2002b}.

\begin{table}
    \centering
    \caption{Character table for point group $C_{6v}$ and the reduction of \vecrep in this group.}
    \begin{tabular}{c|rrrrrr|r}
        \hline\hline
        $C_{6v }$ & $\identity $ & $2C_6$ & $2C_3$ & $C_2$ & $3\sigma_v$ & $3\sigma_d$        \\ \hline
        $A_{1  }$ & $ 1$         & $ 1 $  & $ 1 $  & $ 1$  & $ 1$        & $ 1$               \\
        $A_{2  }$ & $ 1$         & $ 1 $  & $ 1 $  & $ 1$  & $-1$        & $-1$               \\
        $B_{1  }$ & $ 1$         & $-1 $  & $ 1 $  & $-1$  & $ 1$        & $-1$               \\
        $B_{2  }$ & $ 1$         & $-1 $  & $ 1 $  & $-1$  & $-1$        & $ 1$               \\
        $E_{1  }$ & $ 2$         & $ 1 $  & $-1 $  & $-2$  & $0 $        & $0 $               \\
        $E_{2  }$ & $ 2$         & $-1 $  & $-1 $  & $ 2$  & $0 $        & $0 $               \\ \hline
        $\vecrep$ & $ 2$         & $ 1 $  & $-1 $  & $-2$  & $0 $        & $0 $       & $\equiv E_1$\\
        \hline\hline
    \end{tabular}
    \label{tab:character-table-c6v}
\end{table}

\section{Concluding Remarks}
\label{sec:conclusion}
In this work, the origin of the continuous ground-state degeneracy in classical dipolar-coupled systems was traced back to general properties of the underlying lattice. Using the representation theory for the point symmetry group, a generic rule for the degeneracy of Luttinger-Tisza ground states was determined. In doing so, previously known results~\cite{Belobrov1983,Prakash1990,Rastelli2002b,Chamati2014,Chamati2016} could be recovered.
%
In particular, we showed that the ground-state degeneracy of dipolar-coupled LT systems crucially depends on the vector representation \vecrep of the point symmetry group. If the representation \vecrep is irreducible, as for the examples given in Section~\ref{sec:examples}, then a continuous ground-state manifold is found. In contrast, if \vecrep is reducible, a reduced dimension of the degenerate manifold or the absence of a continuous degeneracy altogether is expected.

As the degeneracy only arises in the ground state and is not protected by a symmetry in the Hamiltonian, it is not expected to persist after introducing excitations. We instead expect, in analogy to Ref.~\cite{Prakash1990}, that the inclusion of positional disorder or thermal fluctuations restores the finite symmetry of the Hamiltonian through an order-by-disorder transition~\cite{Belobrov1983,Prakash1990,DeBell1997,Baek2011,Fernandez2007,Carbognani2000,Patchedjiev2005,LeBlanc2006,DeBell1997,Schildknecht2018a,Way2018}. Similarly, higher-order multipoles, especially relevant for artificial spin ice systems, have been found to affect the ground-state degeneracy~\cite{Klymenko1993,Vedmedenko2008}. However, to answer the question of how excitations and disorder affect the ground-state degeneracy, fluctuations on top of a generic system would need to be considered. While this is beyond the scope of the present work, a symmetry-guided discussion of the fluctuations seems feasible.

Finally, we have only considered systems where the LT method is applicable. While this method is valid for many systems~\cite{Luttinger1946,Belobrov1983,Prakash1990}, there exist a number of interesting systems where the LT method does not apply. One example is the system of dipolar-coupled Heisenberg spins on the ``fcc-kagome'' lattice, where a continuous ground-state degeneracy is found~\cite{Way2018}. The ground-state manifold found in Ref.~\cite{Way2018} is not equivalent to a sphere and the basis states are not orthogonal, which is why the LT method does not apply. While such phenomena lie outside the work presented here, it seems feasible to perform a symmetry-guided discussion of non-LT systems, and we hope that this work serves as an inspiration to extend the symmetry discussion to all such systems.

\begin{acknowledgments}
D.S.\ acknowledges partial funding from PSI-CROSS (Grant No. 03.15) and M.S.\ received funding from the European Union’s Horizon 2020 research and innovation program under the Marie Sk\l{}odowska-Curie Grant No. 701647.
\end{acknowledgments}
\bibliography{library}
\end{document}